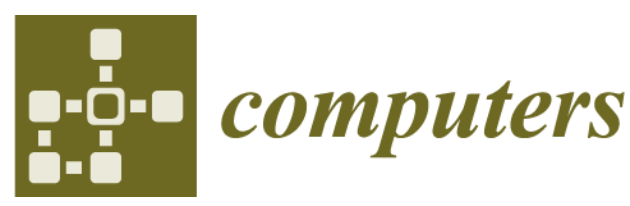
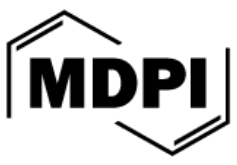

*Article*

# Ecosystem-Driven Privacy Exposure in Mobile Gaming Apps: A Configuration-Aware Empirical Analysis

**Bakheet Aljedaani** [1,*]

[1] Computer Science Department, Jamoum University College, Umm Al-Qura University, Makkah, Saudi Arabia
* Correspondence: bhjedaani@uqu.edu.sa

**Abstract**

Mobile gaming applications (apps) increasingly rely on third-party Software Development Kits (SDKs) for advertising, analytics, attribution, and user engagement, potentially introducing privacy exposure beyond traditional permission-based risks. Existing studies have largely focused on permissions or isolated tracking behaviors, providing only a partial understanding of privacy exposure in modern mobile ecosystems. This study presents a configuration-aware empirical assessment of privacy exposure in Android mobile gaming apps by examining permissions, manifest-level configurations, exported components, and SDK ecosystem complexity across children-oriented and general-audience games. A systematic static analysis was conducted on 41 widely deployed Android mobile gaming apps collected from the Google Play ecosystem. The analysis incorporated SDK categorization and statistical evaluation using Spearman correlation, Mann–Whitney U, and Chi-square testing. The results revealed that privacy exposure is strongly associated with ecosystem-level architectural decisions rather than permission requests alone. Children-oriented games frequently demonstrated exposure conditions comparable to general-audience apps despite sometimes requesting fewer sensitive permissions. Furthermore, larger and more diverse SDK ecosystems were significantly associated with elevated privacy exposure levels, while advertising-oriented SDKs showed strong association with high exposure classifications. These findings highlight the limitations of permission-centric assessment approaches and emphasize the importance of configuration-aware and ecosystem-aware privacy evaluation methodologies for modern mobile software systems.

## 1. Introduction

Mobile gaming applications (apps) represent one of the most widely used categories of mobile software systems, supported by a rapidly expanding global market that was valued at approximately USD 121 billion in 2025 and is projected to exceed USD 289 billion by 2034 [1]. In fact, mobile gaming now represents the largest segment of the broader gaming industry, accounting for nearly half of global gaming revenues and substantially surpassing console and PC gaming segments [2]. Recent industry reports indicate that mobile games generated more than 50 billion downloads globally in 2025 alone, reflecting the scale and ubiquity of mobile gaming ecosystems [3]. In particular, children-oriented mobile games have experienced substantial growth due to the increasing accessibility of smartphones and tablets and the popularity of educational and entertainment-based digital content [4,5]. Modern mobile games increasingly integrate advertising, analytics, attribution, cloud, and social-media-related infrastructures through third-party Software Development Kits (SDKs) to support monetization and user engagement [6,7]. Although these integrations provide important operational benefits, they may simultaneously introduce privacy and security exposure conditions through extensive third-party dependencies, embedded tracking infrastructures, and configuration-level implementation decisions [8,9]. Privacy and security concerns in mobile ecosystems have received growing attention from researchers and practitioners. Prior studies have shown that mobile apps frequently request sensitive permissions, communicate with external tracking domains, and integrate third-party infrastructures capable of collecting behavioral and device-level information [6,10-12]. Earlier information-flow studies further demonstrated that sensitive data may be transmitted through complex application behaviors and external communication pathways not immediately visible through permission declarations alone [13]. Ref. [14] further demonstrated that privacy-relevant data leakage may occur even when Android permission systems appear restrictive, highlighting limitations of permission-centric privacy evaluation. More recent studies suggested that privacy exposure often extends beyond permission requests and may emerge from architectural configurations, software dependencies, and SDK ecosystem complexity [7,15,16].

Privacy concerns become particularly important in children-oriented apps. Previous studies have reported that child-directed mobile apps frequently integrate advertising and analytics ecosystems despite privacy regulations and marketplace policies intended to protect younger users [17,18]. At the same time, recent research investigating marketplace privacy disclosures has shown inconsistencies between developer-reported privacy information and observable app behaviors [8-10,19,20]. These findings suggest that privacy exposure in mobile ecosystems may not be fully represented through permissions or marketplace disclosures alone. Despite significant advances in mobile privacy research, several important gaps remain. First, many existing studies primarily focus on permissions, tracking behaviors, or isolated SDK observations while providing comparatively limited attention to configuration-level exposure indicators and broader ecosystem complexity. Second, although children-oriented apps have received increasing research attention, fewer studies have comparatively examined whether children-oriented mobile games demonstrated lower architectural privacy exposure than general-audience games. Third, prior work has rarely investigated the relationship between SDK ecosystem complexity and privacy exposure levels using a unified software engineering perspective that combines permissions, manifest-level configurations, exported components, and third-party ecosystem analysis within a single empirical framework.

To address these gaps, this study presents a configuration-aware empirical assessment of privacy exposure in Android mobile gaming applications. A dataset of 41 widely deployed Android mobile gaming apps, including both children-oriented and general-audience games, was systematically analyzed using static analysis techniques incorporating permission inspection, manifest-level configuration analysis, exported component analysis, and third-party SDK ecosystem assessment. The study is guided by the following Research Questions (RQs):

*RQ1: What types of privacy exposure indicators are observable in children-oriented and general-audience mobile gaming apps?*

*RQ2: How prevalent are third-party SDK ecosystems in mobile gaming apps, and how do they differ between children-oriented and general-audience games?*

*RQ3: Is there a relationship between SDK ecosystem complexity and privacy exposure levels in mobile gaming apps?*

*RQ4: Do children-oriented apps demonstrate lower SDK ecosystem complexity or lower privacy exposure than general-audience games?*

The findings demonstrate that privacy exposure in mobile gaming apps is strongly influenced by ecosystem-level architectural decisions rather than by permission requests alone. The results further show that children-oriented apps frequently integrate advertising, analytics, and external SDK ecosystems at levels comparable to general-audience games despite sometimes requesting fewer sensitive permissions. This study contributes to the growing body of empirical software engineering and mobile privacy research through the following contributions:

- A configuration-aware empirical assessment framework that combines permission analysis, manifest-level configuration inspection, exported component analysis, and SDK ecosystem assessment for evaluating privacy exposure in mobile applications.
- A comparative analysis of children-oriented and general-audience mobile gaming apps examining differences in architectural exposure conditions and third-party ecosystem integration.
- An empirical investigation of SDK ecosystem complexity and privacy exposure, demonstrating statistically significant relationships between SDK count, SDK diversity, and elevated exposure levels.
- A software engineering perspective on ecosystem-driven privacy exposure that highlights the limitations of permission-centric assessment approaches and emphasizes the role of architectural configurations and third-party dependencies in shaping privacy exposure in modern mobile software systems.

The remainder of this paper is organized as follows. Section 2 presents the related work. Section 3 describes the research methodology. Section 4 presents the results and statistical analysis results. Section 5 discusses the findings. Section 6 reports threats to validity. Finally, Section 7 concludes the paper and outlines future research directions.

## 2. Related Work

Research on privacy and security in mobile ecosystems has expanded significantly over the last decade, particularly in relation to Android applications, third-party tracking infrastructures, advertising ecosystems, and marketplace transparency mechanisms. Existing studies have investigated permission usage, SDK integration, mobile tracking behaviors, privacy disclosures, and security misconfigurations across various application domains. However, comparatively fewer studies have examined privacy exposure in mobile gaming ecosystems using a configuration-aware software engineering perspective that combines permissions, architectural configurations, and SDK ecosystem complexity within a unified empirical framework.

### *2.1 Privacy and Security Analysis of Android Apps*

Several studies have investigated privacy and security exposures in Android apps using static, dynamic, or hybrid analysis techniques. Early research primarily focused on Android permission systems and over-privileged apps. Refs. [11,21,22], for example, demonstrated that many Android apps request permissions beyond their functional requirements, raising concerns regarding excessive privilege access. Subsequent studies expanded the analysis beyond permission by investigating broader architectural and implementation-level issues. Refs. [20,23,24] reported that insecure configurations, implementation behaviors, backend communication practices, and third-party interactions frequently contribute to privacy and security exposures in Android apps. Ref. [12] further demonstrated that personally identifiable information leakage often occurs through network communication patterns and embedded third-party infrastructures rather than through permissions alone.

Although these studies provide important insights into Android privacy and security, many primarily focus on isolated permission analysis, runtime tracking behaviors, or domain-specific apps such as healthcare or banking apps. Comparatively fewer studies have examined mobile gaming ecosystems using a broader configuration-aware perspective incorporating SDK ecosystem complexity and architectural exposure indicators.

### *2.2 Third-Party SDK Ecosystems and Mobile Tracking Infrastructures*

Third-party SDK ecosystems have become a central component of modern mobile apps [7]. Advertising, analytics, attribution, and engagement frameworks are commonly integrated to support monetization and user interaction functionalities. However, prior research has shown that these ecosystems may significantly influence mobile privacy exposure [25].

Ref. [6] demonstrated that advertising libraries frequently introduce additional permissions, hidden functionality, and external communication pathways into Android apps. Ref. [25] similarly reported that advertising libraries commonly collect device identifiers and behavioral information through embedded third-party infrastructures. Refs. [8,26,27] further showed that mobile apps often communicate with multiple external tracking domains capable of enabling large-scale behavioral profiling. Other studies have examined the architectural implications of SDK integration. Refs. [28,29] reported that third-party SDKs frequently inherit broad privileges from host applications, potentially expanding access to sensitive resources. Ref. [7] similarly highlighted how advertising and attribution ecosystems facilitate cross-app tracking through interconnected SDK infrastructures. Refs. [24,30] further suggested that dependency-heavy mobile architectures may introduce undocumented behaviors, hidden data flows, and broader privacy exposure conditions. Related network-level studies have further shown that mobile applications may leak personally identifiable information through externally controlled communication pathways [12].

Collectively, these studies indicate that privacy risks in mobile applications are increasingly shaped by external ecosystem dependencies rather than by application functionality alone. Nevertheless, limited research has comparatively examined SDK ecosystem complexity and architectural exposure conditions specifically within children-oriented and general-audience mobile gaming apps. Privacy analyses focusing specifically on gaming ecosystems remain comparatively limited despite evidence that monetization practices and advertising infrastructures may significantly shape exposure conditions [31].

### *2.3 Privacy Risks in Children-Oriented Mobile Apps*

Children-oriented mobile apps have attracted increasing attention due to concerns regarding behavioral tracking, advertising infrastructures, and data collection practices involving minors [32]. Several studies have reported that child-directed apps frequently embed third-party tracking ecosystems despite marketplace policies and privacy regulations intended to protect younger users [33]. Ref. [17] identified widespread transmission of persistent identifiers and integration of third-party tracking services in children-oriented Android apps. Ref. [34] similarly observed that many children-oriented apps embedded advertising and analytics infrastructures capable of collecting behavioral and device-level information. Ref. [8] further reported that child-directed applications frequently communicate with advertising and analytics services despite presenting themselves as educational or family-oriented apps. Research on mobile tracking ecosystems has also highlighted how combinations of analytics, attribution, and advertising frameworks may facilitate behavioral profiling across apps. Ref. [26] emphasized that these ecosystem dependencies increasingly shape privacy risks within modern mobile environments. Ref. [35] emphasized that network-level investigations have likewise reported extensive third-party communications and advertising-related traffic within popular children-oriented apps.

These findings suggest that children-oriented applications may not necessarily adopt more privacy-preserving architectures than general-audience apps. However, prior work has rarely comparatively evaluated SDK ecosystem complexity and architectural exposure conditions between children-oriented and general-audience gaming applications using a unified empirical assessment framework.

### *2.4 Privacy Transparency, Marketplace Disclosures, and Data Safety Labels*

Another growing area of research focuses on privacy transparency mechanisms in mobile app marketplaces. Recent studies have examined whether privacy labels and developer-reported disclosures accurately reflect observable app behaviors. Ref. [36] found that privacy disclosures are often incomplete or inconsistent with actual data collection practices. Ref. [19] similarly reported that users frequently struggle to interpret privacy and security labels effectively. Refs. [9,20] demonstrated inconsistencies between developer-reported privacy practices and observable technical indicators extracted from apps .

Although transparency-oriented research provides valuable insights into marketplace disclosure mechanisms, most studies focus primarily on privacy labels or behavioral tracking analysis rather than broader architectural exposure conditions. In contrast, the present study adopts a configuration-aware perspective that combines SDK ecosystem analysis, configuration inspection, and exposure-level assessment to investigate privacy exposure in mobile gaming ecosystems.

### *2.5 Research Gap and Positioning of the Present Study*

The reviewed literature demonstrates substantial progress in understanding mobile privacy risks, third-party tracking infrastructures, SDK ecosystems, and marketplace transparency mechanisms. Prior studies have shown that:

- Permissions alone provide only a partial representation of privacy exposure.
- Third-party SDK ecosystems frequently introduce additional tracking and communication pathways.
- Children-oriented apps may still integrate advertising and analytics infrastructures.
- Marketplace privacy disclosures may not accurately reflect observable application behaviors.

However, several important research gaps remain. First, comparatively fewer studies have examined mobile gaming ecosystems using a unified software engineering perspective that combines permission analysis, configuration inspection, exported component analysis, and SDK ecosystem complexity assessment within a single empirical framework. Second, limited research has comparatively evaluated architectural exposure conditions between children-oriented and general-audience mobile gaming apps. Third, prior work has rarely statistically investigated the relationship between SDK ecosystem complexity and architectural privacy exposure levels.

To address these gaps, the present study introduces a configuration-aware empirical assessment framework that combines multiple architectural and ecosystem-level indicators to analyze privacy exposure in Android mobile gaming applications. Unlike prior studies focusing primarily on permissions or isolated tracking behaviors, this work investigates how SDK ecosystem complexity, configuration-level implementation decisions, and third-party dependencies collectively shape exposure conditions in modern mobile gaming ecosystems.

## 3. Research Methodology

This study employed a configuration-aware empirical methodology to investigate privacy exposure in Android mobile gaming apps. The methodology combines APK-based static analysis, SDK ecosystem inspection, and statistical evaluation to examine how permissions, configuration-level properties, and third-party dependencies collectively influence privacy exposure in mobile gaming ecosystems. The study focuses on both children-oriented and general-audience games to support comparative analysis across app categories. Figure 1 presents an overview of the research process. The research process consisted of six main phases: 3.1 Study Design Protocol, 3.2 App Selection and APK Collection, 3.3 Static Analysis and Privacy Indicator Extraction, and 3.4 SDK Ecosystem Analysis, 3.5 Privacy Exposure Classification, and 3.6 Statistical Analysis.

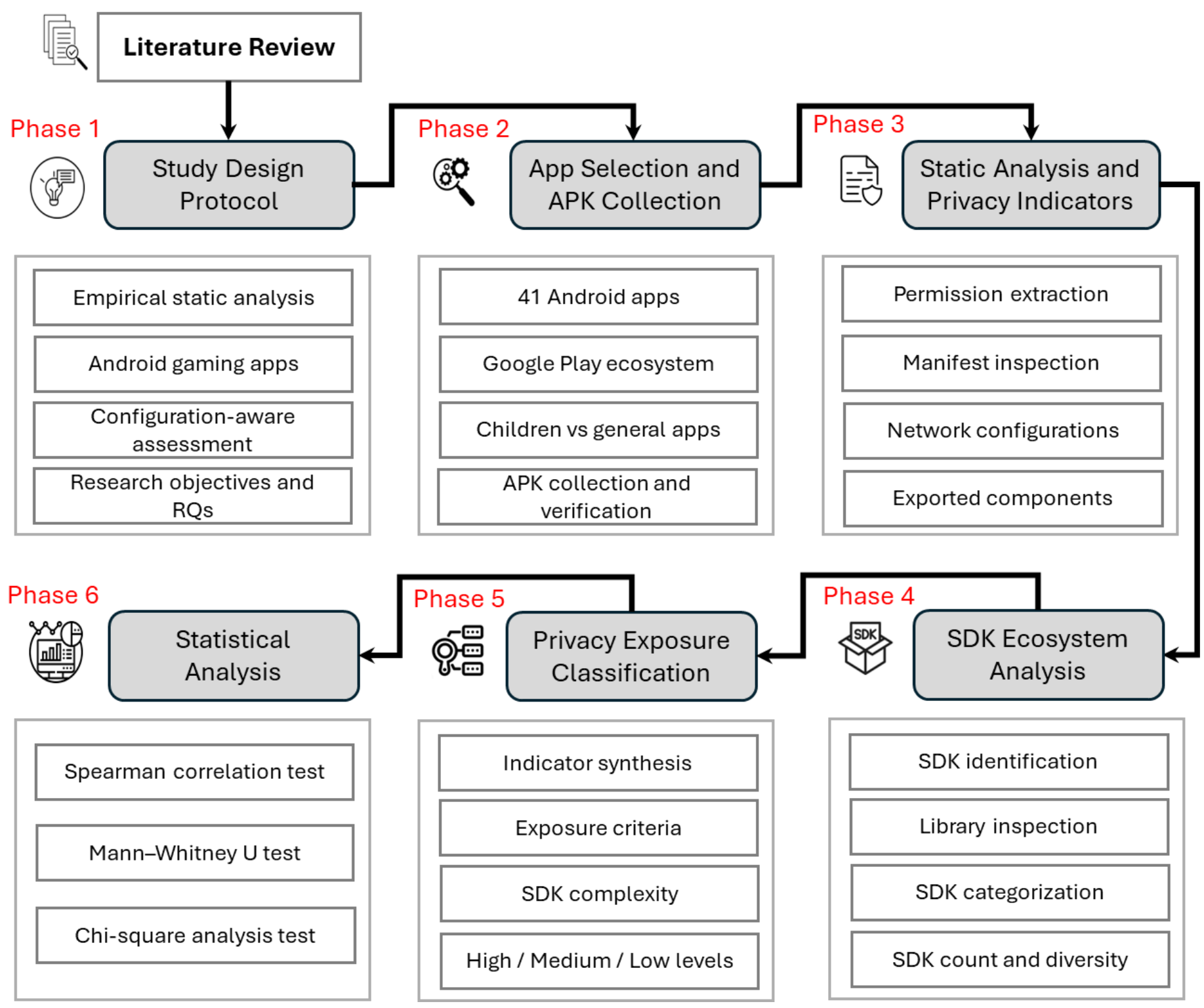

**Figure 1.** An Overview of the Research Process

### *3.1 Study Design Protocol*

The study was designed as an empirical static analysis investigation of Android mobile gaming apps. The primary objective was to examine privacy exposure beyond traditional permission-centric approaches by incorporating configuration-level indicators and SDK ecosystem characteristics into the analysis process. The investigation focused on four aspects: observable privacy exposure indicators, prevalence of third-party SDK ecosystems, relationships between SDK ecosystem complexity and privacy exposure, and comparative differences between children-oriented and general-audience apps.

### *3.2 App Selection and APK Collection*

The dataset consists of 41 Android mobile gaming apps collected from the Google Play ecosystem, including both children-oriented and general-audience games. The selected apps represent widely deployed and actively maintained mobile games with substantial marketplace visibility. Apps were selected using purposive sampling to ensure diversity in popularity, gameplay style, and monetization characteristics. Children-oriented apps were identified using marketplace indicators such as educational positioning and family-oriented classifications, while general-audience games included widely downloaded entertainment-focused titles. APK files were obtained from publicly accessible APK repositories and verified prior to analysis. APK collection was conducted during the defined data collection period to maintain consistency across analyzed app versions.

APK files for the selected apps were obtained from publicly accessible repositories to support static analysis. In some cases, apps were distributed in split-package formats (e.g., XAPK or APKM), requiring additional processing to reconstruct usable APK files. Several downloaded packages were incomplete or corrupted, resulting in extraction failures or inconsistent file structures. These issues were addressed through re-download and verification procedures using alternative repositories while maintaining consistency across selected app versions.

### *3.3 Static Analysis and Privacy Indicator Extraction*

Static analysis was conducted through APK inspection focusing on manifest-level artifacts, configuration properties, and third-party SDK ecosystems. The analysis examined multiple privacy-related indicators to provide a broader assessment of architectural exposure conditions. The static analysis process employed multiple Android reverse-engineering and inspection tools. APK decoding and resource extraction were conducted using apktool version 2.9.3, while manifest inspection and permission extraction were supported using Android Asset Packaging Tool (AAPT) from Android SDK Build-Tools version 36.1.0. Additional package validation and resource inspection were performed using JADX version 1.5.5. These tools were selected due to their widespread adoption in Android software analysis and their suitability for configuration-aware static inspection.

Collected APK files were decoded to enable inspection of internal resources and manifest configurations. The decoding process extracted readable representations of AndroidManifest.xml and related application resources. In some cases, decoding failures occurred due to packaging inconsistencies or obfuscation techniques, requiring repeated extraction attempts or alternative decoding configurations prior to analysis.

#### 3.3.1 Permission Analysis

The first analysis stage focused on extracting declared Android permissions from APK manifest files. The analysis examined both general network-related permissions and sensitive permissions associated with storage access, media access, location services, microphone usage, device identifiers, and communication capabilities. Permission extraction was performed using APK inspection utilities capable of parsing Android manifest configurations.

#### 3.3.2 Configuration-Level Analysis

To move beyond permission-centric assessment approaches, the study additionally inspected manifest-level configuration properties associated with architectural privacy exposure. Prior research has shown that exported and inter-application communication pathways may unintentionally enlarge application attack surfaces and create unintended information flows [37]. Our analysis focused on indicators such as cleartext traffic allowances, backup-related settings, network security configurations, exported activities, exported services, and externally accessible app components. Particular attention was given to properties including *usesCleartextTraffic*, *allowBackup*, *networkSecurityConfig*, and *exported app components*. These indicators were later incorporated into the overall exposure-level classification process. It should be noted that the analyzed indicators, including insecure communication settings and exposed components, are also recognized within established mobile security guidance and industry threat models [38]. Besides, Ref. [39] showed that insecure transport configurations and Secure Sockets Layer (SSL) implementation weaknesses may expose mobile communications to interception and manipulation risks.

### *3.4 SDK Ecosystem Analysis*

A major component of the study involved identifying and categorizing embedded third-party SDK ecosystems. The analysis examined observable SDK artifacts, library references, manifest entries, and package-level indicators associated with external infrastructures integrated into the apps. SDKs were categorized into multiple functional groups, including advertising, analytics, attribution, social-media, monitoring, and monetization infrastructures. Examples of identified SDKs included *Firebase*, *AppsFlyer*, *Facebook SDK*, *AdMob*, *Unity Ads*, and *IronSource*. To support ecosystem-level analysis, two SDK complexity indicators were defined which were later used to statistically evaluate relationships between SDK ecosystem complexity and privacy exposure levels:

- **SDK Count:** the total number of identifiable SDKs or external libraries integrated within an app.
- **SDK Diversity:** the number of distinct SDK ecosystem categories represented within the app architecture.

### *3.5 Privacy Exposure Classification*

Figure 2 presents the conceptual model underlying the privacy exposure assessment adopted in this study. The model illustrates how privacy exposure in mobile gaming apps may emerge from the interaction of multiple architectural dimensions, including resource access permissions, manifest-level configurations, and third-party SDK ecosystems. Rather than relying solely on permission requests, the proposed configuration-aware perspective considers ecosystem dependencies and implementation-level design decisions as contributing factors influencing architectural privacy exposure. These combined indicators informed the exposure classification process described in this section.

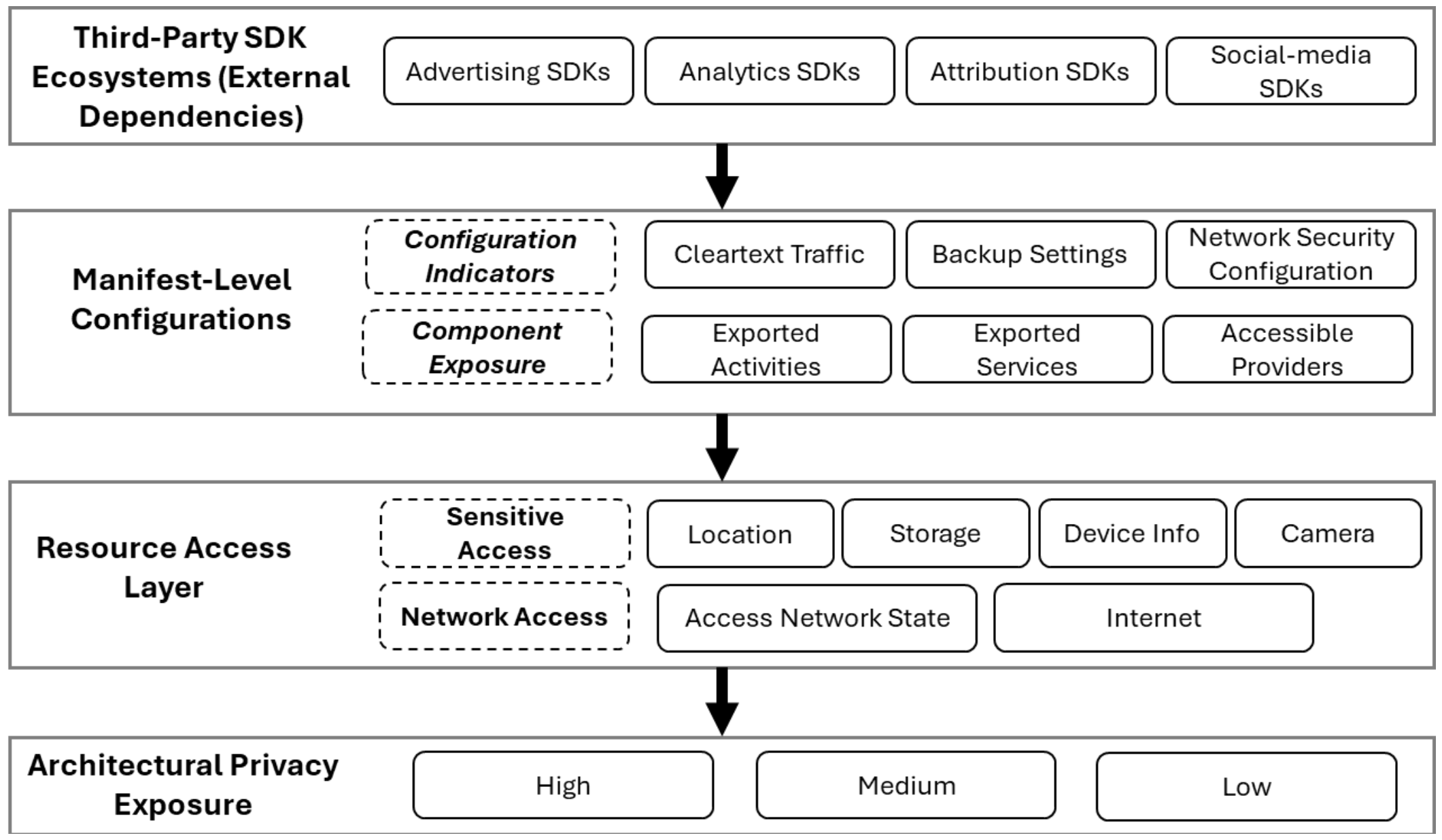

**Figure 2.** Conceptual Model of Architectural Privacy Exposure in Mobile Gaming Apps

To synthesize the extracted indicators, each app was assigned an overall privacy exposure classification based on the combination of observed architectural and ecosystem-level characteristics. The classification incorporated permission usage, sensitive permissions, configuration-level indicators, exported component exposure, and SDK ecosystem complexity. Apps categorized as high exposure typically demonstrated multiple co-occurring indicators, including extensive SDK integration, permissive configurations, numerous exported components, and broad communication capabilities. Medium-exposure apps exhibited partial exposure conditions, while low-exposure apps demonstrated comparatively conservative configurations and limited ecosystem dependencies. The classification process was applied consistently across all analyzed apps using the same evaluation criteria and inspection workflow. Table 1 presents a summary of privacy exposure classification criteria that has been utilized in this study.

**Table 1.** Privacy Exposure Classification Criteria.

| Exposure Level | Typical Indicators |
| --- | --- |
| High | Multiple SDKs, exported components, cleartext traffic |
| Medium | Moderate SDK integration, limited config issues |
| Low | Minimal SDKs, conservative configurations |

*3.6 Statistical Analysis*

Statistical analysis was conducted to investigate relationships between SDK ecosystem characteristics and privacy exposure conditions. Because several variables exhibited non-parametric characteristics, non-parametric statistical techniques were employed. The analysis included:

- **Spearman correlation analysis** to examine relationships between SDK ecosystem complexity and privacy exposure levels,
- **Mann–Whitney U testing** to compare SDK ecosystem characteristics between children-oriented and general-audience apps.
- **Chi-square analysis** to evaluate associations between advertising SDK presence and elevated exposure classifications.

Data organization and statistical analysis were conducted using Microsoft Excel. These techniques enabled comparative evaluation of ecosystem-level characteristics and architectural privacy exposure indicators across the analyzed dataset.

**Ethical Considerations**. The study relied exclusively on the static inspection of publicly distributed Android APKs without interacting with user accounts, personal data, or runtime application behaviors. The analysis focused solely on application-level artifacts and configuration characteristics. To minimize potential harm and avoid unintended reputational impacts, findings are reported in aggregate form and individual apps are anonymized.

## 4. Results

*4.1 Overview of Privacy Exposure Across Mobile Gaming Apps*

The static analysis revealed that privacy exposure in mobile gaming apps extends beyond permission requests and is influenced by multiple architectural and ecosystem-related factors. Across the analyzed dataset, the apps demonstrated varying combinations of sensitive permissions, exported components, permissive network configurations, backup-related settings, and third-party SDK integrations. To synthesize the observations, each app was assigned an overall privacy exposure level based on the combination of indicators identified during the static inspection. The classification incorporated permission usage, manifest-level configurations, exported component exposure, and SDK ecosystem integration. Apps categorized as high exposure typically demonstrated extensive SDK integration, permissive configurations, and broader component exposure surfaces, while medium- and low-exposure apps exhibited comparatively fewer indicators. Among the 41 analyzed apps, 28 apps (68.3%) were classified as high exposure, 12 apps (29.3%) as medium exposure, and 1 app (2.4%) as low exposure. Several children-oriented apps demonstrated exposure characteristics comparable to general-audience games, suggesting that audience designation alone does not necessarily correspond to lower privacy exposure.

### *4.2 SDK Ecosystem Prevalence in Mobile Gaming Apps*

The analysis identified widespread integration of third-party SDK ecosystems across both children-oriented and general-audience mobile games. Multiple categories of SDKs were observed, including advertising, analytics, attribution, social media, monitoring, and monetization frameworks. Table 2 summarizes the prevalence of identified SDKs and libraries across the analyzed apps. Advertising-oriented infrastructures represented the most dominant category, followed by social-media and analytics integrations. Attribution frameworks such as *AppsFlyer* and engagement-related services such as *Helpshift* were also observed. Advertising SDKs were detected in 29 of the 41 analyzed apps (70.7%), while social-media-related SDKs appeared in 25 apps (61.0%). Analytics infrastructures, primarily associated with Firebase ecosystems, were identified in 11 apps (26.8%). Attribution-related SDKs appeared in 7 apps (17.1%). These findings indicate that modern mobile gaming apps frequently rely on extensive third-party infrastructures beyond their core application functionality.

**Table 2.** Prevalence of Third-Party SDK Ecosystems Across the Analyzed Apps.

| SDK Category | Count | Percentage |
| --- | --- | --- |
| Advertising | 29 | 70.7% |
| Social Media | 25 | 61.0% |
| Analytics | 11 | 26.8% |
| Attribution | 7 | 17.1% |

### *4.3 Comparative SDK Integration Between Children-Oriented and General-Audience Apps*

A comparative analysis was conducted to examine differences in SDK ecosystem integration between children-oriented and general-audience apps. The results showed that both categories extensively relied on third-party infrastructures, although some differences in ecosystem composition were observed, as in Table 3. Advertising SDKs were identified in 16 children-oriented apps and 13 general-audience apps. Analytics-related infrastructures appeared more frequently in children-oriented apps, with 8 observed integrations compared to 3 in general-audience games. In contrast, attribution-oriented SDKs were more prevalent among general-audience apps, appearing in 5 apps compared to 2 children-oriented games. Similarly, social-media-related SDKs were substantially more common in general-audience games, where they were identified in 17 apps compared to 8 children-oriented apps. Despite these ecosystem-level differences, substantial SDK integration was observed across both app categories. These findings suggest that children-oriented apps are not isolated from broader tracking and monetization ecosystems and may still exhibit elevated privacy exposure conditions through third-party dependencies.

**Table 3.** Comparative SDK Ecosystem Integration Between Children-Oriented and General-Audience Apps.

| SDK Category | Children-Oriented Apps (n=21) | General-Audience Apps (n=20) | Total |
|---|---|---|---|
| Advertising | 16 | 13 | 29 |
| Analytics | 8 | 3 | 11 |
| Social Media | 8 | 17 | 25 |
| Attribution | 2 | 5 | 7 |

*4.4 Relationship Between SDK Ecosystem Complexity and Privacy Exposure*

To investigate the relationship between SDK ecosystem complexity and privacy exposure, the study analyzed both SDK count and SDK diversity across the exposure categories. SDK count represents the number of distinct SDKs identified within an app, while SDK diversity reflects the number of functional SDK categories integrated into the application ecosystem. As in Figure 3, the results demonstrated a clear increase in both SDK count and SDK diversity as exposure levels increased. Apps classified as high exposure exhibited an average SDK count of 3.50 and an average SDK diversity of 3.07. In contrast, medium-exposure apps demonstrated substantially lower averages, with 1.92 SDKs and 1.75 SDK categories on average. The only low-exposure app exhibited no observable third-party SDK integration. Spearman correlation analysis revealed statistically significant positive associations between SDK count and privacy exposure level ($\varrho = 0.523$, $p = 0.0004$), as well as between SDK diversity and exposure level ($\varrho = 0.505$, $p = 0.0008$). These findings indicate that apps integrating larger and more diverse SDK ecosystems tend to exhibit elevated exposure conditions.

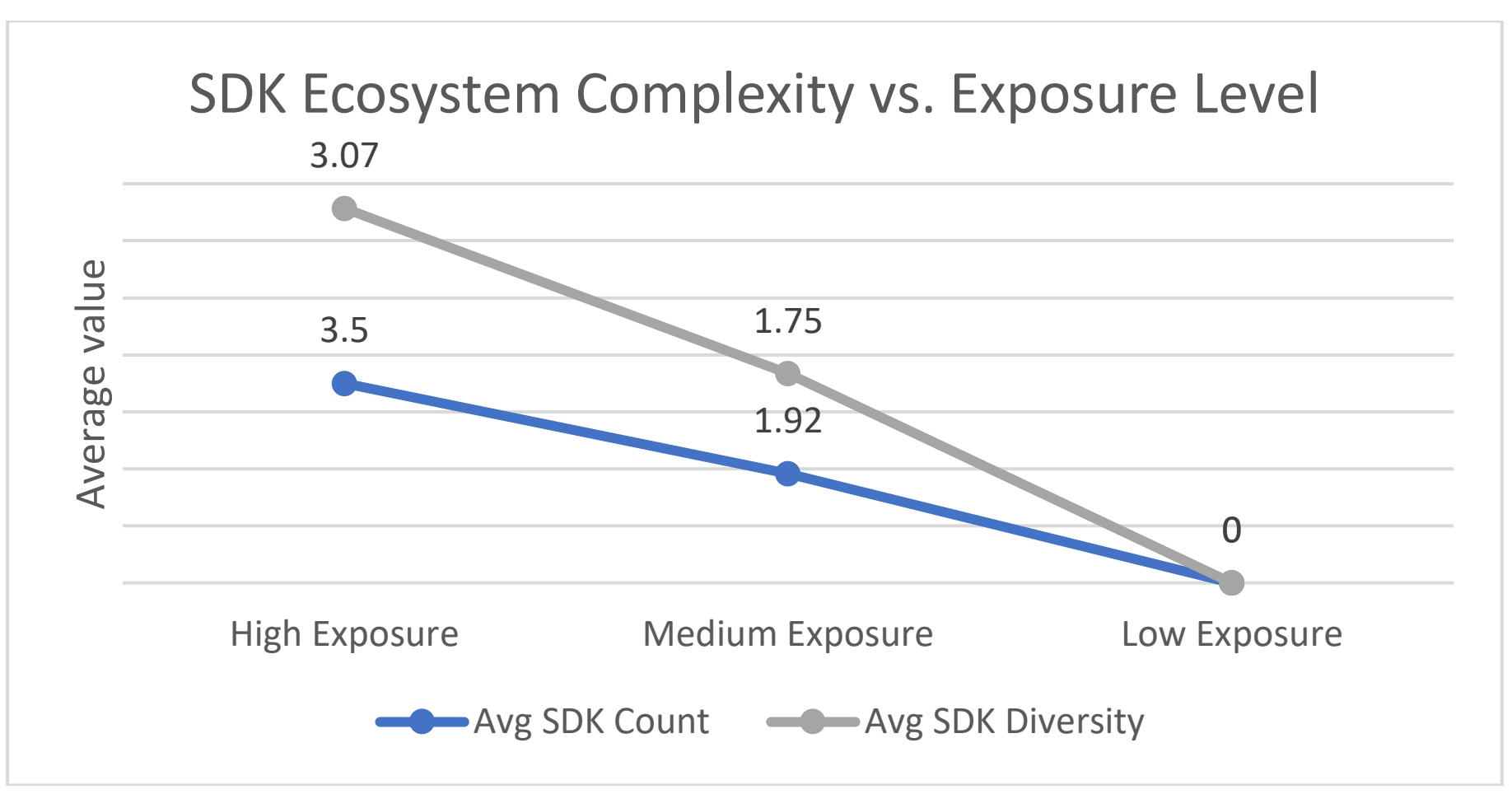

**Figure 3.** Relationship Between SDK Ecosystem Complexity and Privacy Exposure Levels

*4.5 Statistical Comparison of Children-Oriented and General-Audience Apps*

To examine whether SDK ecosystem complexity differed significantly between children-oriented and general-audience apps, a Mann–Whitney U test was conducted using SDK count as the dependent variable. The analysis revealed no statistically significant difference between the two app categories ($U = 218.0$, $p = 0.8406$). This finding suggests that children-oriented apps may integrate third-party SDK ecosystems at levels comparable to general-audience games. Although children-oriented apps sometimes requested fewer sensitive permissions, they still frequently incorporated advertising, analytics, and external ecosystem components.

*4.6 Advertising SDKs and Elevated Privacy Exposure*

The study further examined whether advertising-oriented SDK infrastructures were associated with elevated privacy exposure conditions. A Chi-square test of independence was performed to evaluate the relationship between advertising SDK presence and high-exposure classification. The analysis revealed a statistically significant association between advertising SDK integration and elevated exposure levels ($\chi^2 = 3.952$, $p = 0.0468$). Apps integrating advertising infrastructures were more likely to exhibit broader exposure indicators, including exported components, permissive network configurations, and additional third-party dependencies. These findings suggest that advertising-oriented infrastructures represent a major contributor to privacy exposure conditions in mobile gaming ecosystems. Table 4. provides a summary for the conducted statistical tests in this study.

**Table 4.** Statistical Relationships Between SDK Ecosystem Complexity and Privacy Exposure

| Test | Variables | Statistics | p-value |
|---|---|---|---|
| Spearman | SDK Count vs Exposure | $\rho = 0.523$ | 0.0004 |
| Spearman | SDK Diversity vs Exposure | $\rho = 0.505$ | 0.0008 |
| Mann–Whitney U | Children vs General | U = 218 | 0.8406 |
| Chi-square | Ads SDK vs High Exposure | $\chi^2 = 3.952$ | 0.0468 |

## 5. Discussion

*5.1 Privacy Exposure Extends Beyond Permission Requests*

One of the key findings of this study is that privacy exposure in mobile gaming apps cannot be accurately characterized through permission analysis alone. Although permissions remain an important indicator of app capabilities, the results demonstrate that manifest configurations, exported components, network settings, and third-party SDK integrations substantially contribute to broader exposure conditions. Foundational Android permission research has also demonstrated that permission models contain complex mappings and privilege relationships that may not always be visible to end users or developers [40]. Several apps with moderate permission footprints still exhibited elevated exposure levels due to permissive configurations and extensive external dependencies.

These findings align with prior studies showing that permission requests provide only a partial representation of privacy and security exposures in mobile ecosystems. Ref. [23] identified insecure coding practices, excessive communication with third-party domains, and configuration weaknesses in Android health apps. Similarly, Ref. [24] reported that architectural misconfigurations and insecure backend communication represented major security concerns in mobile banking apps. Ref. [20] further demonstrated that privacy risks frequently emerge from implementation behaviors and ecosystem interactions that are not fully observable through permissions alone.

The findings are also consistent with broader research emphasizing the limitations of permission-centric privacy assessment. Ref. [6] showed that advertising libraries frequently introduce additional communication pathways and externally controlled functionality into Android apps. Ref. [12] similarly demonstrated that personally identifiable information leakage often occurs through embedded third-party components and network interactions. Overall, the results suggest that modern mobile privacy exposures are increasingly shaped by architectural configurations and third-party ecosystem dependencies rather than by permission declarations alone. Consequently, configuration-aware assessment approaches are necessary for evaluating privacy exposure in contemporary mobile apps.

*5.2 Third-Party SDK Ecosystems as a Major Source of Privacy Exposure*

The analysis revealed extensive integration of third-party SDK ecosystems across both children-oriented and general-audience mobile gaming apps. Advertising, analytics, attribution, monitoring, and social-media-related infrastructures were widely embedded throughout the analyzed apps, introducing additional external dependencies and communication pathways. Advertising-oriented infrastructures represented the most dominant ecosystem category identified in the dataset. More than two-thirds of the analyzed apps integrated advertising-related SDKs, while social-media-related integrations appeared in over half of the examined apps. Statistical analysis further revealed a significant association between advertising SDK presence and elevated exposure classifications.

These observations are consistent with previous research highlighting the privacy implications of embedded third-party ecosystems in Android apps. Ref. [6] reported that advertising libraries frequently expand the attack surface of mobile apps by introducing additional permissions and external communication mechanisms. Ref. [25] similarly found that many Android advertising libraries collect device identifiers and behavioral information with limited transparency. Ref. [8] further demonstrated that mobile apps commonly integrate multiple trackers communicating with external domains.

Research on software ecosystems has also shown that dependency-heavy architectures may increase privacy and security exposures in mobile systems. Ref. [28] observed that third-party SDKs frequently inherit broad privileges within Android apps, while Ref. [26] demonstrated that advertising and attribution infrastructures often facilitate cross-app tracking and behavioral profiling. The present study extends this body of work by quantitatively demonstrating that SDK ecosystem complexity itself is significantly associated with elevated exposure levels. This finding suggests that modern mobile privacy risks are increasingly influenced by ecosystem dependencies and monetization infrastructures rather than by core application functionality alone.

*5.3 Children-Oriented Apps Are Not Necessarily More Privacy Preserving*

A particularly important finding of this study is that children-oriented mobile gaming apps do not necessarily demonstrate lower exposure levels than general-audience games. Although several children-oriented apps requested fewer sensitive permissions, many still integrated extensive advertising, analytics, and third-party tracking ecosystems. In addition, the statistical analysis identified no significant difference in SDK counts between children-oriented and general-audience apps. These findings challenge the assumption that apps targeting children inherently adopt more conservative privacy practices. Instead, the results suggest that monetization infrastructures and external ecosystem dependencies remain prevalent even within apps marketed as educational or child-friendly environments.

This observation aligns with prior research examining children-oriented mobile ecosystems. Ref. [17], in a large-scale study of COPPA compliance, found that many child-directed Android apps transmitted persistent identifiers and interacted with third-party tracking services despite regulatory restrictions. Ref. [18] similarly reported that numerous children-oriented apps embedded advertising and analytics frameworks capable of collecting behavioral and device-level information.

Research investigating mobile privacy disclosures has also identified inconsistencies between marketplace privacy representations and observable technical behaviors. Refs. [9,36] found that privacy labels frequently fail to accurately represent real app data practices, while Ref. [19] reported that users often struggle to interpret privacy and security disclosures effectively. Overall, the findings reinforce concerns regarding the adequacy of existing privacy governance approaches for children-oriented mobile ecosystems. The results further suggest that educational positioning or age classification alone should not be interpreted as reliable indicators of privacy-preserving architectural design.

*5.4 SDK Ecosystem Complexity as an Indicator of Architectural Privacy Exposure*

The statistical analysis revealed significant positive correlations between both SDK count and SDK diversity with exposure levels. Apps classified as high exposure consistently demonstrated larger and more diverse third-party ecosystems than medium- or low-exposure apps. These findings indicate that SDK ecosystem complexity itself may serve as an important architectural indicator of privacy exposure in mobile apps. From a software engineering perspective, these observations are particularly significant because modern mobile apps increasingly depend on modular ecosystems composed of analytics, advertising, attribution, cloud, and engagement services. While these integrations provide development efficiency and monetization benefits, they simultaneously expand communication pathways, increase dependency complexity, and enlarge app exposure surfaces.

The observed relationship between SDK ecosystem complexity and exposure levels is consistent with broader software ecosystem research. Ref. [7] reported that third-party SDK ecosystems frequently introduce undocumented behaviors and hidden data flows into Android applications. Ref. [24] similarly showed that dependency-heavy mobile applications often exhibit expanded security and privacy exposures due to inherited permissions and externally controlled behaviors. Prior empirical studies have also demonstrated that SDK ecosystems may contribute to cross-platform tracking and behavioral profiling. Ref. [26] found that mobile tracking infrastructures increasingly rely on combinations of advertising, attribution, and analytics services to construct detailed user interaction profiles. Ref. [8] similarly observed extensive communication between mobile applications and external tracking domains involving multiple third-party entities simultaneously. The present study extends this body of work by quantitatively demonstrating that both SDK count and SDK diversity are significantly associated with elevated exposure levels. This contribution moves beyond binary permission analysis and positions privacy risk as an ecosystem-driven architectural phenomenon shaped by software dependencies and integration complexity.

*5.5 Implications for Privacy Engineering and Mobile Software Development*

The findings of this study have several implications for researchers, developers, and platform providers. First, the results suggest that permission-centric privacy assessment approaches are insufficient for evaluating modern mobile apps. Future privacy engineering methodologies should incorporate configuration analysis, SDK ecosystem inspection, and architectural dependency assessment into both academic and industrial evaluation processes. Second, the widespread integration of advertising and analytics infrastructures within children-oriented apps raises concerns regarding privacy governance and regulatory compliance. Prior studies have repeatedly shown that mobile advertising ecosystems may facilitate persistent tracking and behavioral profiling [6,17,25]. The current findings indicate that such concerns remain relevant even within applications targeting younger audiences. Third, the results highlight limitations in existing privacy transparency mechanisms. Refs. [19,36] suggested that users often cannot accurately assess privacy exposure using marketplace labels or disclosures alone. Even more, Ref. [34] reported persistent gaps between disclosure presentation and users' understanding of underlying data practices The findings of this study reinforce these concerns by showing that substantial exposure conditions may arise from architectural configurations and third-party dependencies that are not always clearly represented in user-facing disclosures. Finally, the study demonstrates the importance of ecosystem-aware empirical software engineering research in the mobile domain. As mobile apps continue evolving toward increasingly interconnected architectures, future privacy and security analyses must account for the influence of third-party services, dependency ecosystems, and configuration-level implementation decisions.

## 6. Threats to Validity

As with any empirical software engineering study, this work is subject to several limitations that should be considered when interpreting the findings. To improve the rigor and transparency of the study, threats to validity are discussed from the perspectives of construct validity, internal validity, external validity, and conclusion validity.

### *6.1 Construct Validity*

Construct validity concerns whether the selected measures accurately represent the investigated concepts. In this study, privacy exposure was inferred from observable static indicators extracted from APK files, including permission usage, manifest-level configurations, exported components, and third-party SDK integrations. Although these indicators are widely used in mobile privacy and security research, they do not necessarily confirm actual malicious behavior or runtime data misuse. Instead, they represent potential exposure conditions inferred from architectural and configuration-level characteristics. In addition, the classification of apps into high, medium, or low exposure categories involved some degree of interpretation. To reduce subjectivity, the same evaluation criteria and inspection workflow were consistently applied across all analyzed apps. The study also combined multiple indicators rather than relying solely on permissions in order to provide a broader assessment of privacy exposure. Another potential limitation relates to SDK categorization, third-party SDKs were classified into functional groups such as advertising, analytics, attribution, and social media based on observable characteristics and prior literature. However, some SDKs may provide overlapping functionalities or evolve over time, potentially affecting categorization accuracy.

### *6.2 Internal Validity*

Internal validity concerns factors that may influence the correctness of the analysis and interpretation process. The study relied primarily on static analysis techniques, including manifest inspection and SDK identification using APK-level artifacts. Consequently, runtime behaviors, dynamically loaded code, encrypted payloads, and server-side interactions may not have been fully captured. To mitigate this limitation, the analysis focused on architectural and configuration-level indicators that could be consistently extracted across all apps without requiring runtime instrumentation. The use of multiple complementary indicators, including permissions, network configurations, exported components, and SDK ecosystems, also helped reduce reliance on a single analysis dimension. Another limitation relates to SDK detection accuracy, some SDKs may use obfuscation techniques or dynamically retrieved modules that are difficult to identify through static inspection alone. To reduce this issue, the analysis combined manifest inspection with library-level observations and cross-verification of identifiable SDK artifacts whenever possible.

### *6.3 External Validity*

External validity concerns the generalizability of the findings beyond the analyzed dataset. The study examined 41 Android mobile gaming apps collected from the Google Play ecosystem, including both children-oriented and general-audience apps. Although the dataset included widely deployed and highly popular games, the findings may not generalize to all categories of mobile apps, smaller app ecosystems, or platforms beyond

Android. In addition, mobile ecosystems evolve rapidly, and app configurations, SDK integrations, and marketplace disclosures may change over time due to app updates, policy changes, or evolving monetization strategies. Consequently, the observed exposure conditions represent a snapshot of the analyzed apps during the data collection period.

Another potential limitation concerns the classification of children-oriented apps, although consistent inclusion criteria and marketplace indicators were used, some apps may target multiple audience groups simultaneously, potentially affecting categorical distinctions between children-oriented and general-audience apps. Despite these limitations, the dataset still provides meaningful insights into contemporary mobile gaming ecosystems because it includes widely downloaded and actively maintained applications representing diverse gameplay and monetization models.

*6.4 Conclusion Validity*

Conclusion validity concerns whether the statistical analyses adequately support the reported findings. The study employed Spearman correlation, Mann–Whitney U testing, and Chi-square analysis to examine relationships between SDK ecosystem characteristics and privacy exposure indicators. These techniques were selected to accommodate the dataset size and the non-parametric nature of several variables. Nevertheless, the relatively moderate dataset size may limit the statistical power for detecting weaker relationships. In addition, correlation-based findings should not be interpreted as direct causal relationships. While the results indicate significant associations between SDK ecosystem complexity and elevated privacy exposure, they do not establish that SDK integration alone directly causes privacy exposure. Finally, some observed indicators may reflect legitimate functional or monetization requirements rather than inherently harmful behavior. Therefore, the findings should be interpreted as indicators of potential privacy exposure rather than definitive evidence of malicious or non-compliant app behavior.

## 7. Conclusions and Future Work

Mobile gaming apps increasingly rely on third-party ecosystems to support advertising, analytics, attribution, monetization, and user engagement functionalities. While these integrations provide important operational benefits, they may also introduce privacy exposure conditions that extend beyond traditional permission-based assessments. This study presented a configuration-aware empirical analysis of Android mobile gaming apps to investigate how SDK ecosystem complexity, architectural configurations, and third-party dependencies influence privacy exposure across children-oriented and general-audience games. Using static analysis, SDK ecosystem inspection, and statistical evaluation, the study combined permission analysis with manifest-level configuration inspection, exported component analysis, and third-party SDK categorization to provide a broader software engineering perspective on privacy exposure in mobile gaming ecosystems.

The findings revealed that advertising- and analytics-related SDK infrastructures were highly prevalent across the analyzed apps, demonstrating extensive reliance on third-party ecosystems. Children-oriented games frequently exhibited privacy exposure conditions comparable to general-audience apps despite sometimes requesting fewer sensitive permissions. In addition, statistical analysis demonstrated significant positive associations between SDK ecosystem complexity and elevated privacy exposure levels, indicating that apps integrating larger and more diverse SDK ecosystems tend to exhibit broader exposure conditions. The results further showed that advertising-oriented SDK infrastructures were significantly associated with high privacy exposure classifications.

Overall, the findings suggest that privacy exposure in modern mobile apps is increasingly shaped by architectural configurations and third-party ecosystem dependencies rather than by permission requests alone. Consequently, permission-centric assessment approaches may provide only a partial representation of privacy exposure in contemporary mobile ecosystems. This study also highlights several practical implications. Developers may benefit from stronger privacy-by-design practices when integrating third-party SDK ecosystems, particularly in applications targeting children. Similarly, marketplace providers may consider incorporating ecosystem-aware validation mechanisms capable of analyzing SDK dependencies and configuration-level indicators during app review processes. More broadly, researchers and practitioners may benefit from adopting multi-dimensional privacy assessment approaches that combine permission analysis with architectural and ecosystem-level inspection techniques.

Future work may extend this study through dynamic analysis techniques capable of observing runtime SDK interactions and network communication behaviors. Additional large-scale studies across other app categories and mobile platforms may also provide broader insights into ecosystem-driven privacy exposure in modern mobile software systems. Overall, this study contributes to the growing body of empirical software engineering and mobile privacy research by demonstrating that privacy exposure in mobile applications emerges from the interaction of configuration-level implementation decisions and third-party ecosystem dependencies.

**Funding:** This research received no external funding.

**Data Availability Statement:** The list of mobile gaming apps used in our study with the full results for each app can be found in the following link: https://shorturl.at/ROfxF

**Acknowledgments:** During the preparation of this manuscript/study, the author used GPT for the purposes of proofreading. The author has taken full responsibility for the content of this study.

**Conflicts of Interest:** The authors declare no conflicts of interest.

## Abbreviations

The following abbreviations are used in this manuscript:

| | |
|---|---|
| Apps | Applications |
| SDK | Software Development Kit |
| RQ | Research Question |
| APK | Android Application Package |
| AAPT | Android Asset Packaging Tool |
| SSL | Secure Sockets Layer |